\documentclass[prl,superscriptaddress,twocolumn]
{revtex4}
\usepackage{graphicx}
\newcommand{\ket}[1]{\mbox{$| #1 \rangle$}}  

\begin{document}

\title{\bf \large NMR Quantum Calculations of the Jones Polynomial\\(Presented in 
part at the ENC, Daytona Beach, FL, U.S.A., 22 April  to 27 April 2007)}

\author{Raimund Marx}
\affiliation{Department of Chemistry, Technische Universit\"at M\"unchen, Lichtenbergstr. 4, D-85747 Garching, Germany}

\author{Amr Fahmy\footnote{To whom correspondence should be addressed: amr\_fahmy@hms.harvard.edu}}
\affiliation{Biological Chemistry and Molecular Pharmacology, Harvard Medical School, 240 Longwood Avenue, Boston, MA 02115, USA}

\author{Louis Kauffman}
\affiliation{University of Illinois at Chicago, 851 S. Morgan Street, Chicago, IL 60607-7045, USA}

\author {Samuel Lomonaco}
\affiliation{University of Maryland Baltimore County, 1000 Hilltop Circle, Baltimore, MD 21250, USA}

\author{Andreas Sp\"orl}
\author{Nikolas Pomplun}
\affiliation{Department of Chemistry, Technische Universit\"at M\"unchen, Lichtenbergstr. 4, D-85747 Garching, Germany}

\author{Thomas Schulte-Herbr\"uggen}

\affiliation{Department of Chemistry, Technische Universit\"at M\"unchen, Lichtenbergstr. 4, D-85747 Garching, Germany}

\author{John Myers}
\affiliation{Gordon McKay Laboratory, Harvard University, 29 Oxford Street, Cambridge, MA 02138, USA}

\author{ Steffen J. Glaser}
\affiliation{Department of Chemistry, Technische Universit\"at M\"unchen, Lichtenbergstr. 4, D-85747 Garching, Germany}

\maketitle

\centerline{\bf Abstract}

The repertoire of problems theoretically solvable by a quantum computer recently expanded to include the approximate evaluation of knot invariants, specifically the Jones polynomial.  The experimental implementation of this evaluation, however, involves many known experimental challenges.  Here we present  experimental results for a small-scale approximate evaluation of the Jones Polynomial by nuclear-magnetic resonance (NMR), in addition we show how to escape from the limitations of NMR approaches that employ pseudo pure states.

Specifically, we use two spin 1/2 nuclei of natural abundance chloroform and apply a sequence of unitary transforms representing the Trefoil Knot, the Figure Eight Knot and the Borromean Rings. After measuring the state of the molecule in each case, we are able to estimate the value of the Jones Polynomial for each of the knots.

\maketitle

\section{Introduction} 
The Jones polynomial \cite{JO}, a great discovery in knot theory, has recently
become an interesting topic for quantum computing.  In particular, the use
of quantum computing has been discussed for approximately evaluating the
Jones polynomial $V(z)$ at selected values of $z$.  For a knot displayed
as a braid of $n$ strands (specified in terms of a sequence of crossings),
these are the values $z$ of the form $z = \exp(2\pi i/k)$ where $k$ is an
integer in the algorithm of Aharonov, Jones and Landau (AJL) \cite{Ah1}.  In \cite{QCJP,3Strand}
a quantum algorithm is given by Kauffman and Lomonanco (KL) for three-strand braids
that  can be used to evaluate the Jones polynomial at a continous range of angles. Most of the computational
cost of the approximate evaluation is the estimation of the trace of a
unitary matrix. The method of estimation described for the AJL algorithm and the KL algorithm 
requires that the quantum computer
separately obtain an estimate of each of the diagonal elements of the
unitary matrix; then these estimates are summed to yield an estimate of
the trace.

The next section reviews the relation of the Jones polynomial for a braid
to a unitary transformation composed of factors that correspond to braid
crossings, so that the problem of evaluating the Jones polynomial reduces
to the problem of evaluating traces of unitary matrix. 

An exposition
of how the KL algorithm (which we use in this paper) can be regareded as a special case of a generalization of
the AJL algorithm is presented after that. In this sense this paper and its sequels will be about experimental implementation of both the KL  and the AJL quantum algorithms for computing the Jones polynomial.

Following this, we present the method whereby an idealized NMR quantum computer \cite{gradientPPS1,gradientPPS2}
can evaluate the trace of unitary matrix written as a product of factors
all at once, that is, without having to evaluate diagonal elements
of the unitary matrix separately.

Experimental results for the evaluation of cases
of the 2-by-2 matrix, and hence of the Jones polynomial for a braid
of 3 strands, by use of nuclear magnetic resonance (NMR) is the subject of the last section.

\section{The Jones Polynomial and unitary matrices}
\label{JonesPoly}

The Jones polynomial \cite{JO} was a great discovery in knot 
theory. It marked the beginning of a significant relationship between 
knot theory
and  statisical mechanics, particularly through the relationship of 
the polynomial with the Temperley-Lieb algebra, and through the 
explicit
bracket state sum model \cite{KA87,KA88,KA89,KP,KL}.
From the topological side the Jones polynonmial is striking because 
it can detect the difference between many knots and their mirror images.
\vspace{3mm}

  The general algorithm to find the Jones polynomial is in the $\sharp 
P$ complexity class, and so this is an algorithm worth understanding in the
context of quantum computation.

The key idea behind the present quantum algorithms to compute the Jones polynomial is to use unitary 
representations of the braid group derived from Temperley-Lieb algebra representations that take the form
$$\rho(\sigma_{i}) = AI + A^{-1}U_{i}$$
where $\sigma_{i}$ is a standard generator of the Artin braid group, $A$ is a complex number of unit length, and
$U_{i}$ is a symmetric real matrix that is part of a representation of the Temperley-Lieb algebra. For more details about this
strategy and the background information about the Jones polynomial, the bracket model for the Jones polynomial and the Temperley-Lieb algebra
the reader may wish to consult \cite{Ah1,JO,KA87,KA88,KA89,KL,KP,QCJP,Fibonacci,3Strand,shor-jordan}. In the following mathematical description, we have
given a minimal exposition of the structure of such representations.

\subsection{Two Projectors and a Unitary Representation of the Three 
Strand Braid Group}
It is useful to think of the Temperley Lieb algebra as generated by projections
$e_{i} = U_{i}/\delta$ so that $e_{i}^{2} = e_{i}$ and $e_{i}e_{i\pm 
1}e_{i} = \tau e_{i}$ where
$\tau = \delta^{-2}$ and $e_{i}$ and $e_{j}$ commute for $|i-j|>1.$
\vspace{3mm}

With this in mind, consider elementary projectors
$e = |A \rangle \langle A|$ and $f=|B \rangle \langle B|$. We assume that $ \langle A|A \rangle = \langle B|B \rangle =1$ so 
that $e^{2} = e$ and $f^{2} =f.$
Now note that

$$efe = |A \rangle \langle A|B \rangle \langle B|A \rangle \langle A| = \langle A|B \rangle \langle B|A \rangle e = \tau e$$

\noindent Thus $$efe=\tau e$$

\noindent where $\tau = \langle A|B \rangle \langle B|A \rangle $.

This algebra of two projectors is the simplest instance of a 
representation of the Temperley Lieb
algebra.  In particular, this means that a representation of the 
three-strand braid group is
naturally associated with the algebra of two projectors.
\vspace{3mm}

Quite specifically if we let $\langle A| = (a,b)$ and $|A \rangle = (a,b)^{T}$ the 
transpose of this row vector,
then

$$e=|A \rangle \langle A| =  \left[
\begin{array}{cc}
      a^{2} & ab  \\
      ab & b^{2}
\end{array}
\right] $$

\noindent is a standard projector matrix when $a^{2} + b^{2} = 1.$ 
To obtain a specific
representation, 
\smallskip

let
$e_{1} =  \left[
\begin{array}{cc}
      1 & 0  \\
      0 & 0
\end{array}
\right] $
and
$e_{2} =  \left[
\begin{array}{cc}
      a^{2} & ab  \\
      ab & b^{2}
\end{array}
\right] .$

\noindent  It is easy to check that
$e_{1}e_{2}e_{1} = a^{2}e_{1}$
and that
$e_{2}e_{1}e_{2} = a^{2}e_{2}.$

\noindent Note also that
$e_{1}e_{2}=  \left[
\begin{array}{cc}
      a^{2} & ab  \\
      0 & 0
\end{array}
\right]$
and
$e_{2}e_{1}=  \left[
\begin{array}{cc}
      a^{2} & 0  \\
      ab & 0
\end{array}
\right].$

We define $$U_{i} = \delta e_{i}$$ \noindent for $i=1,2$ with $a^{2} 
= \delta^{-2}.$
Then we have , for $i = 1,2$
$$U_{i}^{2} = \delta U_{i} \, ,\, U_{1}U_{2}U_{1} = U_{1},\, U_{2}U_{1}U_{2} = U_{2}.$$
Thus we have a representation of the Temperley-Lieb algebra on three strands. See \cite{KL} for a discussion of the properties
of the Temperley-Lieb algebra.

Note also that we have 
$$trace(U_{1})=trace(U_{2}) = \delta,$$
while
$$trace(U_{1}U_{2}) = trace(U_{2}U_{1}) = 1$$
where {\it trace} denotes the usual matrix trace.
We will use these results on the traces of these matrices 
in Section \ref{threeStrands}.
\vspace{3mm}

Now we return to the matrix parameters:
Since $a^{2} + b^{2} = 1$ this means that $\delta^{-2} + b^{2} = 1$ whence
$b^{2} = 1-\delta^{-2}.$

\noindent Therefore $b$ is real when $\delta^{2}$ is greater than or 
equal to $1$.
\vspace{3mm}

We are interested in the case where $\delta = -A^{2} - A^{-2}$ and 
{\em $A$ is a unit complex
number}.  Under these circumstances the braid group representation
\begin{displaymath}
\rho(\sigma_{i}) = AI + A^{-1}U_{i}
\end{displaymath}
\noindent will be unitary whenever $U_{i}$ is a real symmetric 
matrix. Thus we will obtain a
unitary representation of the three-strand braid group $B_{3}$ when 
$\delta^{2} \geq 1$.

\noindent For any $A$ with $d = -A^{2}-A^{-2}$ these formulas define a representation of the braid group. With 
$A=exp(i\theta)$, we have $d = -2cos(2\theta)$. We find a specific range of
angles $\theta$ in the following disjoint union of angular intervals
$$\theta \in [0,\pi/6]\sqcup[\pi/3,2\pi/3]\sqcup[5\pi/6,7\pi/6]\sqcup[4\pi/3,5\pi/3]\sqcup[11\pi/6,2\pi]$$
{\it that give unitary representations of
the three-strand braid group.} Thus a specialization of a more general represention of the braid group gives rise to a continuous family
of unitary representations of the braid group.

\subsection{A Quantum Algorithm for the Jones Polynomial
on Three Strand Braids}

\label{threeStrands}

We gave above an example of a unitary representation of the three-strand 
braid group.In fact, we can use this representation to compute the Jones 
polynomial for closures of 3-braids, and therefore this
representation provides a test case for the corresponding quantum 
computation. We now analyse this case by first making
explicit how the bracket polynomial is computed from this 
representation. This unitary representation and its application to
a quantum algorithm first appeard in \cite{QCJP}. When coupled with 
the Hadamard test, this algorithm gets values for the Jones polynomial
in polynomial time in the same way as the AJL algorithm \cite{Ah1}. 
It remains to be seen how fast these algorithms are in principle when 
asked to compute the polynomial itself rather than certain specializations of it.
\vspace{3mm}

First recall that the representation depends on two matrices $U_{1}$ 
and $U_{2}$ with

$U_{1} =  \left[
\begin{array}{cc}
      \delta & 0  \\
           0 & 0
\end{array}
\right] $
and \,
$U_{2} =  \left[
\begin{array}{cc}
      \delta^{-1} & \sqrt{1-\delta^{-2}}  \\
          \sqrt{1-\delta^{-2}}  & \delta - \delta^{-1}
\end{array}
\right]. $

\noindent The representation is given on the two braid generators by
\begin{equation}
\rho(\sigma_{1})= AI + A^{-1}U_{1}
\end{equation}
and
\begin{equation}
\rho(\sigma_{2})= AI + A^{-1}U_{2}
\end{equation}

\noindent for any $A$ with $\delta = -A^{2}-A^{-2}$, and with
$A = exp(i\theta)$, then $\delta = -2cos(2\theta)$. We get the specific range of
angles $\theta \in [0,\pi/6]\sqcup[\pi/3,2\pi/3]\sqcup[5\pi/6,7\pi/6]\sqcup[4\pi/3,5\pi/3]\sqcup[11\pi/6,2\pi]$ that 
give unitary representations of
the three-strand braid group.
\vspace{3mm}

Note that $tr(U_{1})=tr(U_{2})= \delta$ while $tr(U_{1}U_{2}) = 
tr(U_{2}U_{1}) =1.$
If $b$ is any braid, let $I(b)$ denote the sum of the exponents in 
the braid word that expresses $b$.
For $b$ a three-strand braid, it follows that
$$\rho(b) = A^{I(b)}I + \tau(b)$$
\noindent where $I$ is the $ 2 \times 2$ identity matrix and 
$\tau(b)$ is a sum of products in the Temperley Lieb algebra
involving $U_{1}$ and $U_{2}.$ Since the Temperley Lieb algebra in 
this dimension is generated by $I$,$U_{1}$, $U_{2}$,
$U_{1}U_{2}$ and $U_{2}U_{1}$, it follows that
$$\langle \overline{b} \rangle = A^{I(b)}\delta^{2} + tr(\tau(b))$$
\noindent where $\overline{b}$ denotes the standard braid closure of 
$b$, and the sharp brackets denote the bracket polynomial
as described in previous sections. From this we see at once that
$$ \langle \overline{b} \rangle = tr(\rho(b)) + A^{I(b)}(\delta^{2} -2).$$
It follows from this calculation that the question of computing the 
bracket polynomial for the closure of the three-strand
braid $b$ is mathematically equivalent to the problem of computing 
the trace of the matrix $\rho(b).$

The matrix in question is a product of unitary matrices, the quantum 
gates that we have associated with the braids
$\sigma_{1}$ and $\sigma_{2}.$ The entries of the matrix $\rho(b)$ 
are the results of preparation and detection for the
two dimensional basis of qubits for our machine:
$$ \langle i|\rho(b)|j \rangle.$$
\noindent Given that the computer is prepared in $|j \rangle $, the 
probability of observing it in state $|i \rangle $ is equal
to $| \langle i|\rho(b)|j \rangle|^{2}.$ Thus we can, by running the quantum 
computation repeatedly, estimate the absolute squares of the
entries of the matrix $\rho(b).$  This will not yield the complex 
phase information that is needed for either the trace of the
matrix or the absolute value of that trace.
\bigbreak
However, we do know how to write a quantum algorithm to compute the 
trace of a unitary matrix (via the Hadamard test).
Since $\rho(b)$ is unitary, we can use this approach to approximate 
the trace of $\rho(b).$ This yields a quantum algorithim for
the Jones polynomial for three-stand braids (evaluated at points $A$ 
such that the representation is unitary).
Knowing $tr(\rho(b))$ from the quantum computation, we then have the 
formula for the bracket, as above,
$$ \langle \overline{b} \rangle = trace(\rho(b)) + A^{I(b)}(\delta^{2} -2).$$
Then the normalized polynomial, invariant under all three 
Reidemeister moves is given by
$$f(\overline{b}) = (-A^{3})^{-I(b)} \langle \overline{b} \rangle .$$
Finally the Jones polynomial in its usual form is given by the formula
$$V(\overline{b})(t) = f(\overline{b})(t^{-1/4}).$$

Thus we conclude that our quantum computer can approximate values of 
the Jones polynomial.

\section{On the relationship with the AJL algorithm}
\label{KL-AJL}
 
Here is how the KL (Kauffman-Lomonaco) algorithm described in the previous section
becomes a special case of a generalization of the AJL algorithm:
Here we use notation from the AJL paper. In that paper, the generators
$U_{i}$ (in our previous notation) for the Temperley-Lieb algebra, are denoted by
$E_{i}.$
\bigbreak

Let $L_{k} = \lambda_{k} = sin(k \theta).$ For the time being $\theta$ is an
arbitrary angle. Let $A = iexp(i \theta/2)$ so that
$d = -A^{2} - A^{-2} = 2cos(\theta).$
\bigbreak

We need to choose $\theta$ so that $sin(k \theta)$ is
non-negative for the range of $k$'s we use (these depend on the choice of
line graph as in AJL). And we insist that $sin(k \theta)$ is non-zero except
for $k=0.$ Then it follows from trigonometry that
$(L_{k-1} + L_{k+1})/L_{k} = d$ for all $k.$
\bigbreak

Recall that the representation of the Temperley-Lieb algebra in AJL is
given in terms of $E_{i}$ such that $E_{i}^2 = dE_{i}$ and the $E_{i}$ satisfy
the Temperley-Lieb relations. Each $E_{i}$ acts non-trivially at the $i$ and
$i+1$ places in the bit-string basis for the space and each $E_{i}$ is based
upon $L_{a-1},L_{a},L_{a+1}$ where $a = z(i)$ is the endpoint of a walk
described by the bitstring using only first $(i-1)$ bits.
Bitstrings represent walks on a line graph. Thus $1011$ represents the walk
Right, Left, Right, Right ending at node number $3$ in
$$1-----2-----3-----4.$$
For $p = 1011,$ $z(1) = 1, z(2) = 2, z(3) = 1, z(4) = 1, z(5) = 3.$

More precisely, if we let
$$|v(a) \rangle = [ \sqrt{L_{a-1}/L_{a}}, \sqrt{L_{a+1}/L_{a}} ]^{T}$$
(i.e. this is a column vector. T denotes transpose.) Then
$$E_{i} = |v(z(i)) \rangle \langle v(z(i))|.$$
Here it is understood that this refers to the action
on the bitstrings $$----------01----------$$ and $$----------10----------$$ obtained from
the given bitstring by modifying the $i$ and $i+1$ places. The basis order is
$01$ before $10.$ Conceptually, this is a useful description, but it also helps
to have the specific formulas laid out.

Now
look at the special case of a line graph with three nodes and two edges:
$$1-----2-----3.$$
The only admissible binary sequences are $|110 \rangle $ and $|101 \rangle ,$ so the space
corresponding to this graph is two dimensional, and it is acted on by
$E_{1}$ with $z(1) = 1$ in both cases
(the empty walk  terminates in the first node)
and
$E_{2}$ with $z(2) = 2$ for $|110 \rangle $ and $z(2)=2$ for $|101 \rangle .$
Then we have
$$E_{1}|110 \rangle = 0, E_{1}|101 \rangle = d |101\rangle ,$$
$$E_{2}|xyz \rangle = |v \rangle \langle v|xyz \rangle $$   (xyz = 101 or 110)
where $v = (\sqrt{1/d}, \sqrt{d - 1/d})^{T}.$
\bigbreak

If one compares this two dimensional representation of the three
strand Temperley - Lieb algebra and the corresponding braid group representation, with
the representation Kauffman and Lomonaco use in their paper, it is clear that it is
the same (up to the convenient replacement of $A = exp(i \theta)$ by $A = i exp(i \theta/2)$). The trace formula of AJL is a variation of
the trace formula that Kauffman and Lomonaco use. Note that the AJL algorithm as formulated in \cite{Ah1} does not
use the continuous range of angles that are available to the KL algorithm. In the sequel to this paper
and in a separate paper on the mathematics, we
shall show how the entire AJL algorithm generalizes to continuous angular ranges.
\bigbreak

\section{Theory of an NMR spectrometer used as a quantum computer}
\label{nmrTheory}

By convention, a quantum computer as conceived in theory is assumed to
yield an outcome associated with a quantum measurement of some (possibly
mixed) quantum state.  In contrast, NMR machines implement a restricted
version of an Expectation-Value Quantum Computer (EVQC), which in place of
an outcome yields, to some finite precision, the expectation value for a
measurement of a (again, possibly mixed) quantum state
\cite{gradientPPS2}.  Reflecting facts of NMR spectrometers, an NMR
Quantum Computer (NMRQC) implements only the special measurement
operators discussed in \cite{Thermal_DJ}, and these measurement operators all
have zero trace.

Here are the details.  For a Hermitian measurement operator $M$ applied to
a density matrix $\rho$, the EVQC of precision $\epsilon$ yields a value
$x$ such that
\begin{equation}
 |x - \mbox{Tr}(M\rho)| \le \epsilon \Lambda(M),
\label{eq:eqvc}
\end{equation}
where $\Lambda(M)$ is the difference between the minimum and the maximum
eigenvalue of the measurement operator $M$, which is just the possible
range to the trace as $\rho$ varies over all possible density matrices.
(The factor $\Lambda(M)$ makes limitations of resolution immune to the
mere analytic trick of multiplying the measurement operator by a
constant.)

The measurement operators of main interest for the algorithm by which
we estimate the trace of a unitary operator are $I_{1x}$
and $I_{1y}$, shortly to be defined.

\subsection{Thermal Equilibrium and initial state preparation}
To first order,  the initial thermal state density operator of an ensemble (very large number) of quantum systems with $n+1$ qubits each  \cite{Ernst} is given by
\begin{equation}
\label{Eq_thermal}
\rho_{th} \approx {{1}\over{N}} ({\bf 1} - \sum_{l=1}^{n+1} \alpha_{l} I_{lz})
\end{equation}
with $\alpha_{l} =\frac{\hbar \omega_l}{{\rm k}T}$, 
\begin{displaymath}
I_{lz} = {{1} \over {2} }{\bf 1}  \otimes  \dots \otimes {\bf 1}  \otimes \sigma_{z}  \otimes {\bf 1}  \otimes \dots \otimes {\bf 1},
\end{displaymath}
(where the Pauli matrix $\sigma_z$ appears as the $l^{th}$ term in the product), $\omega_l$ is the resonance frequency of qubit $l$, ${\rm k}$ is Boltzmann's constant, $T$ is temperature and $N = 2^{n+1}$.

The initial density operator required for our algorithm is given by
\begin{equation}
 \rho_{0} = {{1} \over {N}} ({\bf 1} - \alpha_{1} I_{1z})
\end{equation}
which can be prepared from $\rho_{th}$ by a variety of methods
\cite{fahmy}.

\subsection{Algorithm to estimate the trace of $U$}

The method presented here is based on the algorithm that first appeared in \cite{knill}.
As mentioned above we assume that $U$ is given in the form of local operations on $n$ qubits. Given a program for $U$, Barenco et al. \cite{committeepaper} describe a procedure to construct  a program or local operations for the operator controlled-U, $cU$. $cU$ operates on $n+1$ qubits, does not affect the first qubit,   applies $U$ on the remaining $n$ qubits if the first qubit is $\ket{1}$ and does nothing otherwise:
\begin{eqnarray*}
cU |1\rangle \ket{\psi}  & = & ({\bf 1} \otimes U) \ket{1} \ket{\psi}  = \ket{1}  U \ket{\psi} \\
cU \ket{0}\ket{\psi} & = & \ket{0} \ket{\psi}.
\end{eqnarray*} 

In block matrix form, $cU$ is given by:
\begin{displaymath}
cU = \left( \begin{array}{c c}  {\bf 1}& 0 \\
                                     0 & U \\   
         \end{array}\right)
\end{displaymath}

We now describe our algorithm:

{\bf Step 1:} Prepare the  density operator:
\begin{equation}
 \rho_{1} = {{1} \over {N}} ({\bf 1} - \alpha_{1} I_{1x}) =  {{1} \over {N}} {\bf 1} - {{\alpha_1} \over {2N}} \left( \begin{array}{c c}  0& {\bf1} \\
                   {\bf 1} & 0 \\   
         \end{array}\right),
\end{equation}
where 
\begin{displaymath}
I_{1x} = {{1} \over {2} }\sigma_{x}  \otimes  {\bf 1} \mbox{ and }
\sigma_x = \left[\begin{array}{cc}0 & 1\\1 & 0\end{array}\right].
\end{displaymath}

{\bf Step 2:}  Apply $cU$ to $\rho_{1}$:
\begin{equation}
 \rho_{2} = cU \rho_{1} cU^{\dagger} =  {{1} \over {N}} {\bf 1} - {{\alpha_1} \over {2N}} \left( \begin{array}{c c}  0& U^{\dagger} \\
                    U & 0 \\   
         \end{array}\right).
\end{equation}

{\bf Step 3:} Measure $\langle I_{1x}  +iI_{1y}\rangle$ to estimate
\begin{equation}
trace((I_{1x} + iI_{1y})\;\rho_{2}) = {{ \alpha_1} \over {N}} 
trace(U) ,
\end{equation}
where
\begin{displaymath}
I_{1y} = {{1} \over {2} }\sigma_{y} \otimes {\bf 1} \mbox{ and } \sigma_y
= \left[\begin{array}{rr}0 & -i\\i & 0\end{array}\right] .
\end{displaymath}
By Eq.\ (\ref{eq:eqvc}) the result of this measurement 
is a complex number, z,  such that
\begin{eqnarray}
  |\mbox{Re}(z) - \frac{\alpha_1}{N} \mbox{Re}[\mbox{{\it trace}}(U)]| \le \epsilon \nonumber \\
  | \mbox{Im}(z)- \frac{\alpha_1}{N} \mbox{Im}[\mbox{{\it trace}}(U)]| \le \epsilon.
\end{eqnarray}
It follows that the measurement result satisfies
\begin{equation}
  |z - \mbox{{\it trace}}(U)| \le \sqrt{2}N\epsilon/\alpha_1.
\end{equation}

\section{Example knots and experimental results}
\label{ExpResults}

Experimental results for 3 knots on three strands were obtained using the methods outlined above. Specifically, we present results for the Trefoil Knot, the Figure Eight Knot, the Borromean Rings.  A 2-spin system (details of the molecule and pulse sequences are given later) was used, the initial state given by the density operator proportional to $I_{1x}$ was prepared and a reference spectrum was then collected. This was followed by application of a controlled-unitary operator corresponding to and representing each knot separately found from the representation
\begin{displaymath}
s_{1} = \rho(\sigma_1) \;\;\;\; {\rm and} \;\;\;\; s_{2} =   \rho(\sigma_2).
\end{displaymath}
Measurement of the expectation value of $I_{1x}+iI_{1y}$ after applying the controlled-unitary operator yields the trace of the unitary operator representing the knot and thus the estimate of the Jones Polynomial for each knot.

For each of the three knots, the Jones Polynomial was estimated at the complex numbers $e^{i \theta}$ for all $\theta$ in the range $0 \leq \theta \leq \pi/6$ at single degree increments (31 values).
Comparison to the theoretical values shows excellent correspondence with experimental observations. Furthermore, the Jones Polynomial itself for each of these knots can be constructed from the experimental results. 

\subsection{Experimental setup and molecule}
All experiments were performed on a Bruker Avance DMX 750 NMR spectrometer, equipped with a TXI 5mm probe head with XYZ gradients. The sample was a 9:1 mixture of chloroform and deuterated acetone. It naturally contained about 1\% of  $^{13}$C-$^{1}$H chloroform which was the active compound that represented the "hardware" of our NMR quantum computer. The spin system's Larmor frequencies were 188.6349005 MHz for $^{13}$C and 750.1354275 MHz for $^1$H. The corresponding chemical shifts are 77.2 ppm and 7.235 ppm, respectively. The two spin-$^1/_2$ nuclei of $^{13}$C-$^{1}$H chloroform interact through scalar coupling. The corresponding coupling constant is J=209.5 Hz. The longitudinal relaxation times (T$_1$) and transversal relaxation times (T$_2$) of both spin-$^1/_2$ nuclei are: $^{13}$C T$_1$: 21.8 sec, $^{13}$C T$_2$: 0.19 sec, $^{1}$H T$_1$: 6.1 sec, $^{1}$H T$_2$: 0.48 sec. In order to suppress the signal of 99\% $^{12}$C-$^{1}$H chloroform and to prepare the initial operator $I_x$, where "$I$" corresponds to " $^{13}$C" ($^1$H will be referred to as "$S$"), the following preparation sequence was used in all experiments: the $^1$H spins were saturated by cw irradiation. Subsequently they were dephased by applying a 9.9 $\mu$s 90$^\circ$($^1$H) pulse followed by a B$_0$ gradient. This sequence of 90$^\circ$ pulse and gradient was repeated twice with orthogonal gradients. Subsequently the $^{13}$C spin was excited using a 19.45 $\mu$s 90$^\circ$($^{13}$C) pulse. This preparation sequence was followed by the pulse sequence of the individual experiments (see Figure 7). Finally the $^{13}$C signal was detected by measuring 512 points during 452 ms. In order to improve the sensitivity, we decoupled all $^{1}$H spins during the detection period by applying the DIPSI-2 \cite{dipsi} decoupling sequence.

\section{Conclusion}
\label{conclusions}

In this paper, we showed how the KL algorithm is a special case of a generalized AJL algorithm. Using the KL algorithm, we obtained a unitary representation of the three-strand braid group and discussed a method for computing the Jones polynomial using this representation over a range of complex numbers. Next, the theory of an idealized NMR quantum computer was presented and we showed how the trace of a unitary matrix can be experimentally determined. Experimental realization for three different knots where performed where the experimental data agreed with theoretical calculations. Future work includes generalizing the AJL algorithm for any number of strands, as was done for the three-strand braid group in this work, and their experimental implementations.

\section{Acknowledgements}
AF thanks NIH GM47467.
Pictures for knots and links were created using KnotPlot.
http://knotplot.com/

\begin{figure}
\center{\includegraphics[width=2in, keepaspectratio=true]{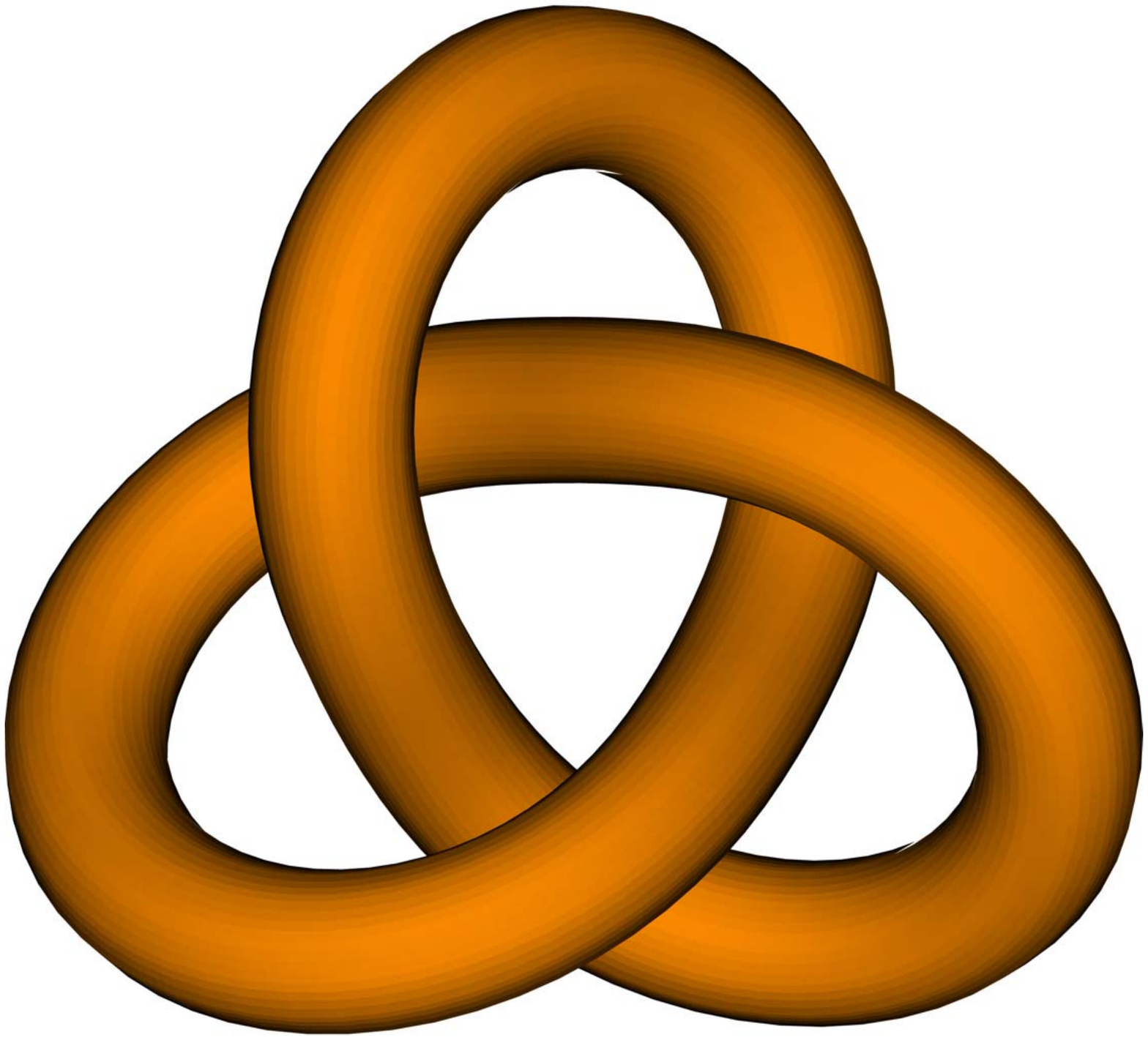}} 
\caption{The Trefoil Knot generated by the sequence $\sigma_1^{3}$}
\end{figure}

\begin{figure}
\center{\includegraphics[width=4in, keepaspectratio=true]{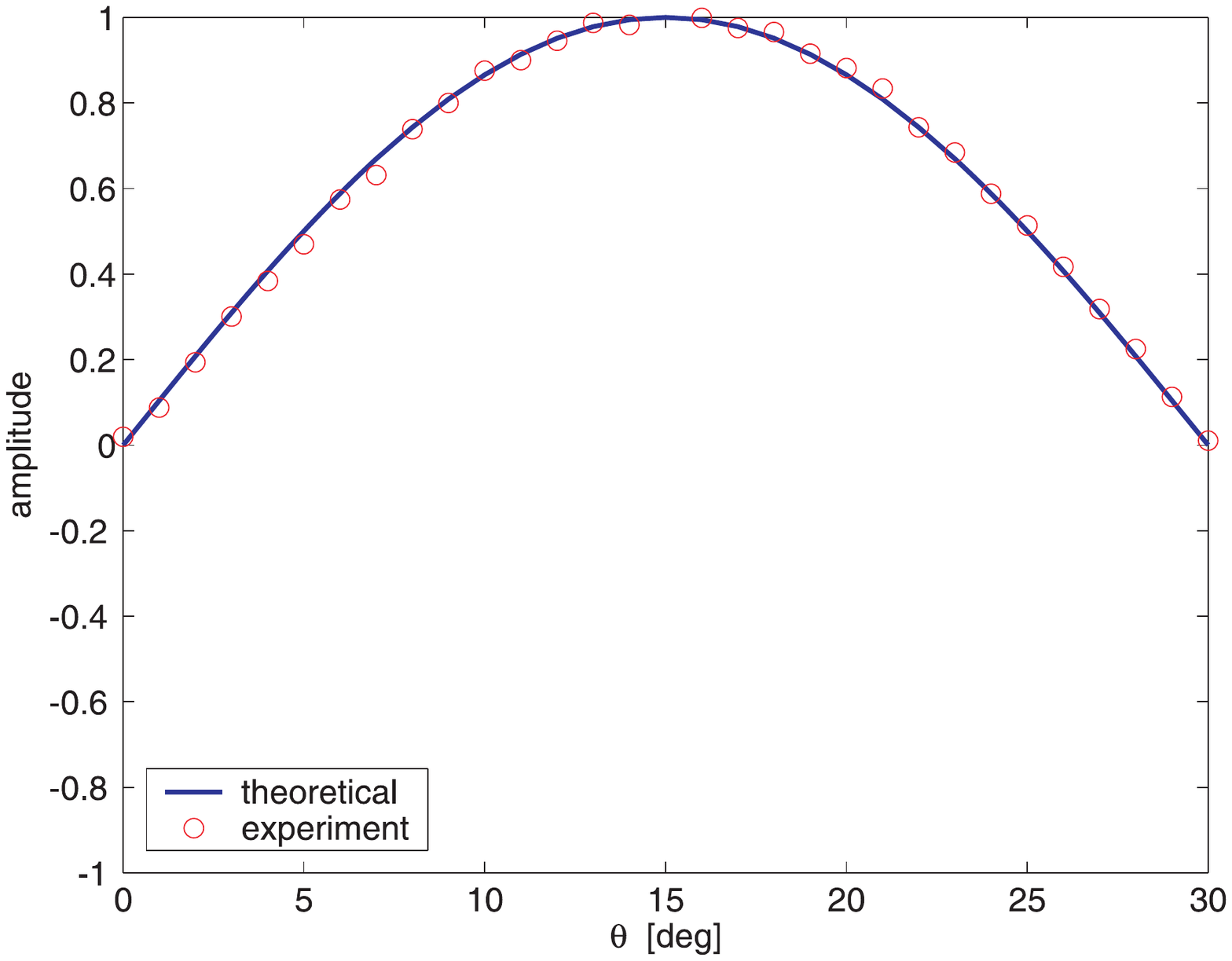}} 
\caption{Experimental results for the Trefoil Knot}
\end{figure}

\begin{figure}
\center{\includegraphics[width=2in, keepaspectratio=true]{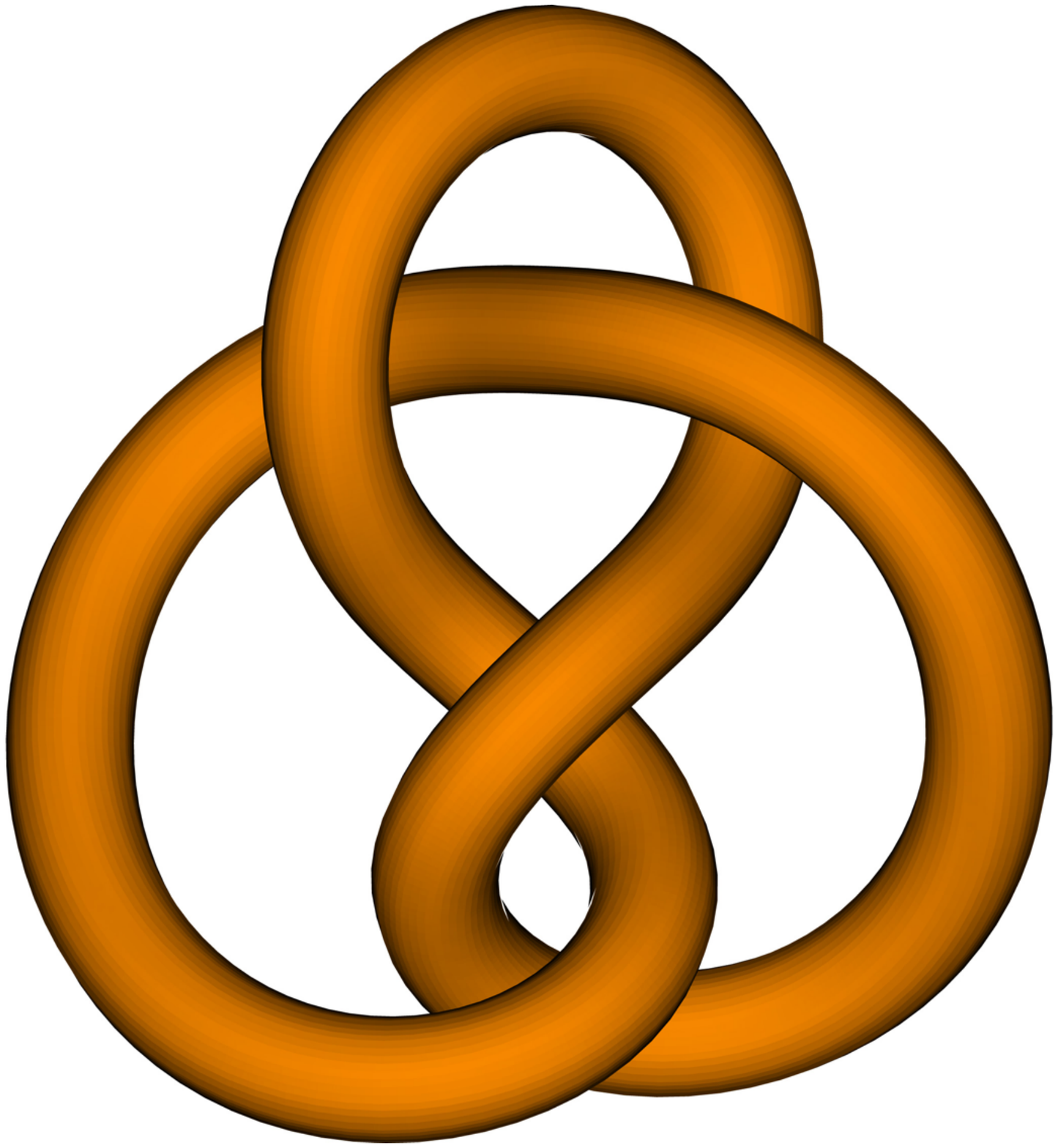}} 
\caption{The Figure Eight Knot generated by the sequence $\sigma_1\;\;\sigma_2^{-1}\sigma_1\;\;\sigma_2^{-1}$}
\end{figure}

\begin{figure}
\center{\includegraphics[width=4in, keepaspectratio=true]{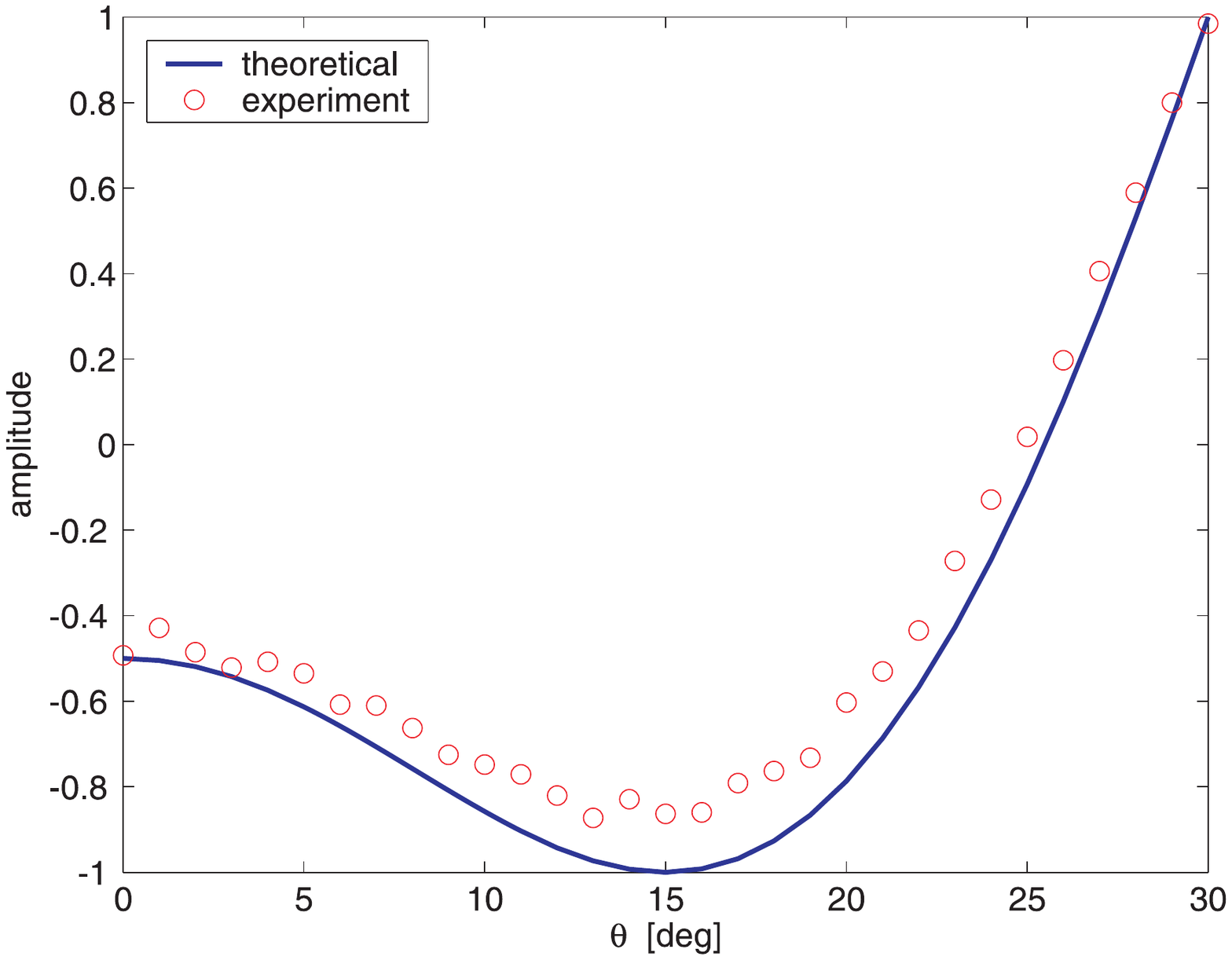}} 
\caption{Experimental results for the Figure Eight Knot}
\end{figure}

\begin{figure}
\center{\includegraphics[width=2in, keepaspectratio=true]{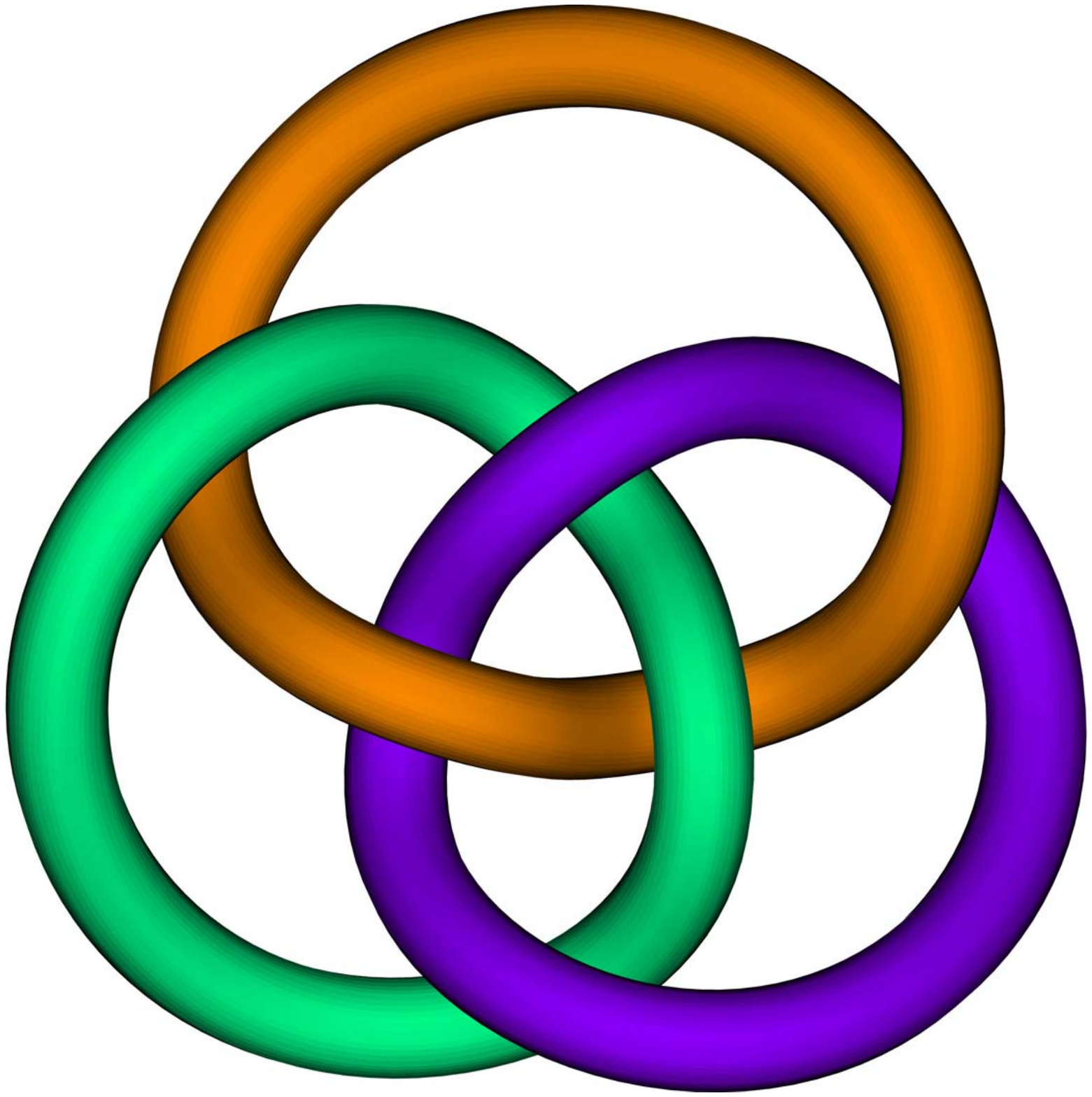}} 
\caption{The Borromean Rings generated by the sequence $\sigma_1 \sigma_2^{-1} \sigma_1 \sigma_2^{-1}  \sigma_1 \sigma_2^{-1}$}
\end{figure}

\begin{figure}
\center{\includegraphics[width=4in, keepaspectratio=true]{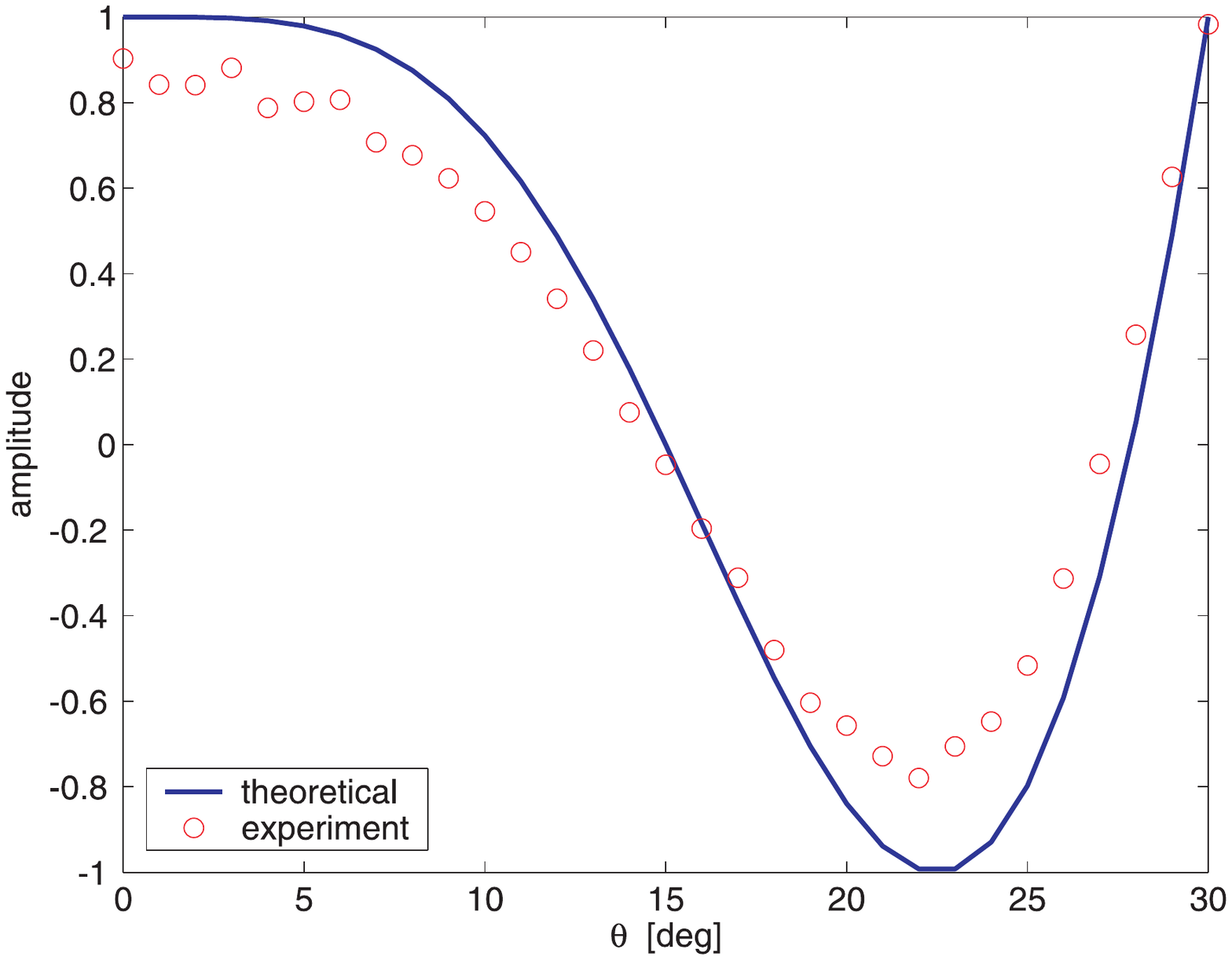}} 
\caption{Experimental results for the Borromean Rings}
\end{figure}

\begin{figure}
\center{\includegraphics[width=4in, keepaspectratio=true]{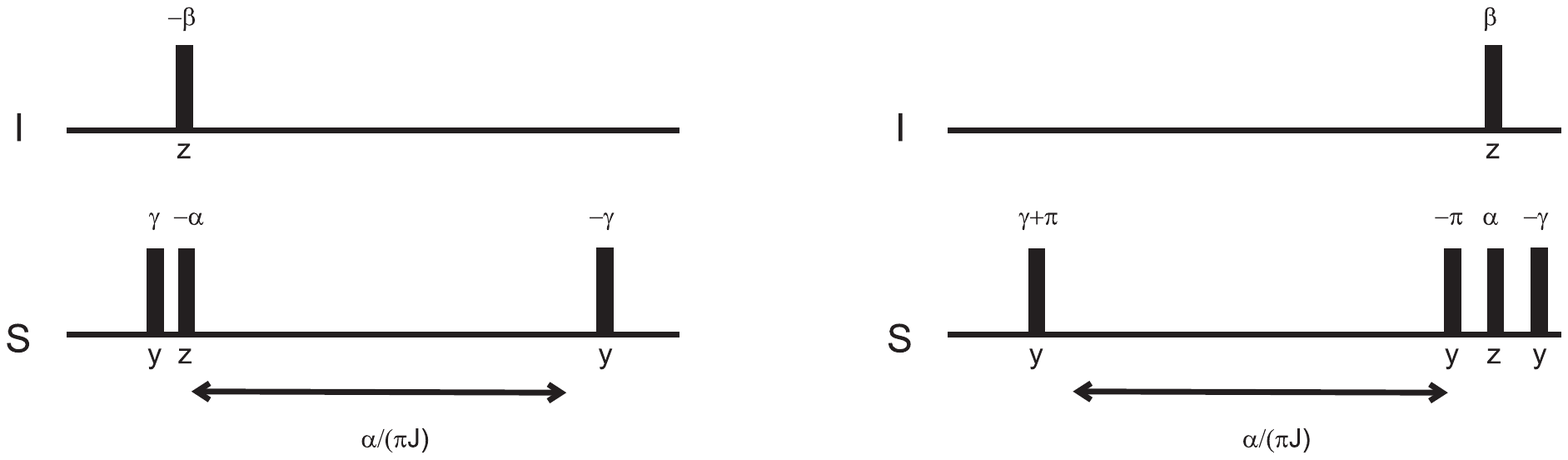}} 
\caption{Pulse sequence to implement a controlled-$s_{1,2}$ operation (left) and its inverse (right). For $s_{1}$ set $\gamma$ to 0. (To obtain a propagator of ${\bf 1} \oplus s_{1,2}$ respectively its inverse, we have to apply a global phase factor of $e^{\mp i \pi\beta/2}$. The propagator of a pulse on the second spin-1/2 is defined as $U_{pulse} := e^{-i\epsilon S_{\nu}}$, with $S_{\nu} := ({\bf 1} \otimes \sigma_{\nu}/2), \epsilon \in \{\alpha,\beta,\gamma\}$ and $\nu \in \{y,z\}$ where $\alpha = 0.5 \pi - 2 \theta$, $\beta =0.5 \pi + \theta$ and $\gamma = \tan^{-1}[\cos{4\theta}/\sqrt{4\cos^{2}{2\theta-1}}]+\pi/2$ and $0 \leq \theta \leq \pi/6$. For the free evolution $U_{evolution} := e^{-i \pi t J 2 I_{z}S_{z}}$ with $2I_{z}S_{z} := 2 (\sigma_{z} /2 \otimes \sigma_{z}/2)$. The pulse sequence is applied to the initial density operator $I_{1x}$}
\end{figure}


\begin{thebibliography}{[28]}

\bibitem{JO} 
V.F.R. Jones, A polynomial invariant for links via von Neumann algebras,
Bull. Amer. Math. Soc. {\bf 129} (1985), 103--112.

\bibitem{Ah1}
D. Aharonov, V. Jones, Z. Landau,
A polynomial quantum algorithm for approximating the Jones polynomial,
quant-ph/0511096.

\bibitem{QCJP}
L.H. Kauffman, Quantum computing and the
Jones polynomial, math.QA/0105255, in {\em Quantum Computation and Information}, S. Lomonaco, 
Jr. (ed.), AMS CONM/305, 2002, pp.~101--137.

\bibitem{3Strand}
Kauffman, Louis H. and Samuel J. Lomonaco, Jr., A 3-Stranded Quantum Algorithm for the Jones Polynomial,
Proc. SPIE, vol. 6573, (2007), 65730T-1-65730T-13. http://arxiv.org/abs/0706.0020

\bibitem{gradientPPS1} D.~G. Cory, A.~F. Fahmy, and T.~F. Havel,
``Nuclear magnetic resonance spectroscopy: An experimentally accessible
paradigm for quantum computing,'' in {\it Proc.\ of the 4th Workshop on
Physics and Computation} (New England Complex Systems Institute,
Boston, MA, 1996), pp.~87--91.

\bibitem{gradientPPS2} D.~G. Cory, A.~F. Fahmy, and T.~F.
Havel, Proc.\ Natl.\ Acad.\ Sci.\ USA {\bf 94}, 1634 (1997).

\bibitem{KA87}  
L.H. Kauffman, State models and the Jones polynomial, Topology {\bf 26} (1987),
395--407.

\bibitem{KA88} L.H.Kauffman, New invariants in the theory of knots,
{\em Amer. Math. Monthly}, Vol.95,No.3,March 1988. pp 195-242.

 \bibitem{KA89}  
L.H. Kauffman,  Statistical mechanics and the Jones polynomial,  AMS
Contemp. Math. Series  {\bf 78} (1989), 263--297.

\bibitem{KL}
L.H. Kauffman, {\em Temperley-Lieb Recoupling Theory and Invariants of Three-Manifolds},
Princeton University Press, Annals Studies {\bf 114} (1994). 

\bibitem {KP}
L.H. Kauffman, {\em Knots and Physics}, World Scientific Publishers (1991), 
Second Edition (1993), Third Edition (2002).

\bibitem{Fibonacci}
L. H. Kauffman and S. J. Lomonaco Jr., The Fibonacci Model and the Temperley-Lieb Algebra.
{\it International J. Modern Phys. B}, Vol. 22, No. 29 (2008), 5065-5080.

 \bibitem{shor-jordan}
P. Shor and S. Jordan. Quant. Inf. and Comm., 8, 681-714, (2008).

\bibitem{Thermal_DJ} 
 J.~M. Myers, A.~F. Fahmy, S.~J. Glaser and R. Marx, 
 Phys.\ Rev. A {\bf  63}, 032302 (2001).

\bibitem{Ernst} R.~R. Ernst, G. Bodenhausen, and A. Wokaun, {\it
Principles of Nuclear Magnetic Resonance in One and Two
Dimensions} (Oxford University Press, Oxford, 1987).

\bibitem{fahmy} 
 A.~F. Fahmy, R. Marx, W. Bermel and S.~J. Glaser, 
 Phys.\ Rev. A {\bf  78}, 022317 (2008).

\bibitem{knill} E.~Knill~and~R.~Laflamme,
Phys. Rev. Lett. 81, 5672-5675 (1998).

\bibitem{committeepaper}
A. Barenco, C.H. Bennett, R. Cleve, D. DiVincenzo, N. Margolus, P. Shor, T. Sleator, J. Smolin, H. Weinfurter, Phys. Rev. A {\bf 52}, 3457 (1995).

\bibitem{dipsi}
A.J. Shaka, C. J. Lee, and A. Pines,
Iterative Schemes for Bibear Operators;
Application to Spin Decoupling, 
J. Magn.\ Reson.\ {\bf 77}, 274 (1988).

\end{thebibliography}
\end{document}